\documentclass[aps,twocolumn,showpacs,superscriptaddress]{revtex4}%
\usepackage{amsfonts}
\usepackage{amsmath}
\usepackage{amssymb}
\usepackage{graphicx}%
\providecommand{\U}[1]{\protect\rule{.1in}{.1in}}

\begin{document}
\title{First-principles LDA+\emph{U} and GGA+\emph{U} study of neptunium dioxide}
\author{Baotian Wang}
\affiliation{Institute of Theoretical Physics and Department of
Physics, Shanxi University, Taiyuan 030006, People's Republic of
China} \affiliation{LCP, Institute of Applied Physics and
Computational Mathematics, Beijing 100088, People's Republic of
China}
\author{Hongliang Shi}
\affiliation{LCP, Institute of Applied Physics and Computational
Mathematics, Beijing 100088, People's Republic of China}
\affiliation{SKLSM, Institute of Semiconductors, Chinese Academy of
Sciences, People's Republic of China}
\author{Weidong Li}
\affiliation{Institute of Theoretical Physics and Department of Physics, Shanxi University,
Taiyuan 030006, People's Republic of China}
\author{Ping Zhang}
\thanks{Author to whom correspondence should be
addressed. E-mail: zhang\_ping@iapcm.ac.cn} \affiliation{LCP,
Institute of Applied Physics and Computational Mathematics, Beijing
100088, People's Republic of China} \affiliation{Center for Applied
Physics and Technology, Peking University, Beijing 100871, People's
Republic of China} 

\begin{abstract}
We have performed a systematic first-principles investigation to calculate the
electronic structures, mechanical properties, and phonon dispersion curves of
NpO$_{2}$. The local density approximation$+U$ and the generalized gradient
approximation$+U$ formalisms have been used to account for the strong on-site
Coulomb repulsion among the localized Np $5f$ electrons. By choosing the
Hubbard \emph{U} parameter around 4 eV, the orbital occupancy characters of Np
5\emph{f} and O 2\emph{p} are in good agreement with recent experiments [J.
Nucl. Mater. \textbf{389}, 470 (2009)]. Comparing with our previous study of
ThO$_{2}$, we note that stronger covalency exists in NpO$_{2}$ due to the more
localization behavior of 5\emph{f} electrons of Np in line with the
localization-delocalization trend exhibited by the actinides series.

\end{abstract}
\maketitle

\section{INTRODUCTION}

Neptunium-based oxides have been extensively investigated by experiments
\cite{Benedict,Yamashita,Serizawa,Santini,Nishi,Seibert} and theoretical
calculations \cite{Kurosaki,Prodan2,Andersson,Petit} due to the technological
importance that neptunium is one of the important long-lived actinides which
accumulate in high level waste during the conventional nuclear fuel cycle.
Among neptunium oxides, NpO$_{2}$ has attracted much more attention because of
its more stable thermodynamical properties and interesting subtle
non-collinear magnetic structure \cite{Serizawa,Yamashita,Benedict,Santini}.
NpO$_{2}$ is reported to can maintain stable under high temperature up to 1400
K \cite{Serizawa}. Seibert \emph{et al.} systematically studied the reaction
of neptunium with molecular and atomic oxygen as well as the the formation and
stability of surface oxides; they reported that high neptunium oxides can only
exist as surface phases not as stable bulk phases \cite{Seibert}. As for the
electronic structures of actinide dioxides (AO$_{2}$), taking (A=U, Np or
Pu)O$_{2}$ for example, the 5\emph{f} orbitals are partially occupied,
accompanying the competition between the quantum process of localization and
delocalization of A 5\emph{f} electrons, which leads to many complex
behaviors. This also makes the accurate description of electronic structure of
AnO$_{2}$ series to be difficult achieved.

Conventional density functional theory (DFT) that apply the local density
approximation (LDA) or generalized gradient approximation (GGA) underestimates
the strong on-site Coulomb repulsion of the 5\emph{f} electrons and,
consequently, describes actinide dioxides as incorrect ferromagnetic (FM)
metals instead of antiferromagnetic (AFM) Mott insulators. One more promising
way to improve the drawback is LDA+\emph{U} or GGA+\emph{U} approach, in which
the underestimation of the strong intraatomic Coulomb interaction is corrected
by the Hubbard \emph{U} parameter. Recently, the electronic structures,
mechanical properties, and high-pressure behaviors of UO$_{2}$ and PuO$_{2}$
have been correctly reproduced using LDA+\emph{U} or GGA+\emph{U} calculations
\cite{DudarevUO2,Geng,Andersson,SunJCP,SunCP,Jomard}. The insulator character
of NpO$_{2}$ is established by the experiments\cite{Veal,Seibert}. However, to
our knowledge, a systematical theoretical investigation of electronic
structure and mechanical properties for NpO$_{2}$ is still lacking.
Consequently, based on the good performance of LDA/GGA+\emph{U} approaches in
describing the electronic structure of the strong-correlation systems, we
carried out the present study of NpO$_{2}$.

In the present study, the lattice parameter, electronic structure, mechanical
features, and thermodynamic properties of NpO$_{2}$ are calculated by
employing the LDA+\emph{U} and GGA+\emph{U} schemes due to Dudarev \emph{et
al}. \cite{Dudarev}. We discuss how the choice of \emph{U} as well as the
choice of exchange-correlation potential, i.e., the LDA or the GGA, affect
those properties. Our results show that the pure LDA or GGA fails to give the
accurate lattice parameter and correct electronic structure, while the
LDA+\emph{U} and GGA+\emph{U} schemes can effectively remedy these failures.
The rest of this paper is arranged as follows. In Sec. II the computational
method is briefly described. In Sec. III we present and discuss our results.
In Sec. IV we summarize the conclusions of this work.

\section{computational methods}

Our total energy calculations are carried out by employing the plane-wave
basis pseudopotential method as implemented in Vienna \textit{ab initio}
simulation package (VASP) \cite{Kresse3}. The exchange and correlation effects
are described by the DFT within LDA and GGA \cite{LDA,GGA}. The projected
augmented wave (PAW) method of Bl\"{o}chl \cite{PAW} is implemented in VASP
with the frozen-core approximation. Electron wave function is expanded in
plane waves up to a cutoff energy of 500 eV, and all atoms are fully relaxed
until the Hellmann-Feynman forces become less than 0.02 eV/\AA . A 9$\times
$9$\times$9 Monkhorst-Pack \cite{Monk} \emph{k} point-mesh in the full wedge
of the Brillouin zone is used for fluorite NpO$_{2}$. The neptunium
6\emph{s}$^{2}$7\emph{s}$^{2}$6\emph{p}$^{6}$6\emph{d}$^{2}$5\emph{f}$^{3}$
and oxygen 2\emph{s}$^{2}$2\emph{p}$^{4}$ electrons are treated as valence
electrons. The strong on-site Coulomb repulsion among the localized Np
5\emph{f} electrons is described by using the formalism formulated by Dudarev
\emph{et al.} \cite{Dudarev}. In this scheme, the total LDA (GGA) energy
functional is of the form
\begin{align}
E_{\mathrm{{LDA(GGA)}+U}}  &  =E_{\mathrm{{LDA(GGA)}}}\nonumber\\
&  +\frac{U-J}{2}\sum_{\sigma}[\mathrm{{Tr}\rho^{\sigma}-{Tr}(\rho^{\sigma
}\rho^{\sigma})],}%
\end{align}
where $\rho^{\sigma}$ is the density matrix of \emph{f} states with spin
$\sigma$, while \emph{U} and \emph{J} are the spherically averaged screened
Coulomb energy and the exchange energy, respectively. In this work, the
Coulomb \emph{U} is treated as one variable, while the parameter \emph{J} is
set to 0.6 eV. Since only the difference between \emph{U} and \emph{J} is
meaningful in Dudarev's approach, therefore, we label them as one single
parameter \emph{U} for simplicity.

\begin{figure}[ptb]
\begin{center}
\includegraphics[width=1.0\linewidth]{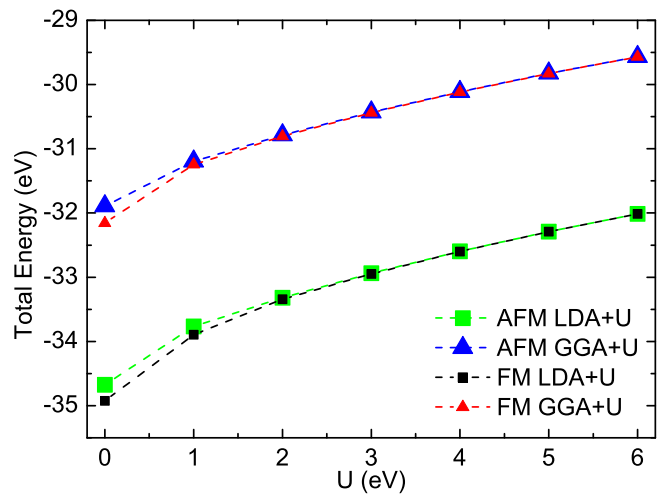}
\end{center}
\caption{(Color online) Dependence of the total energies (per formula unit) on
\emph{U} for FM and AFM NpO$_{2}$.}%
\label{toten}%
\end{figure}

Both spin-unpolarized and spin-polarized calculations are performed in this
study. Compared with FM and AFM phases, the nonmagnetic (NM) phase is not
energetically favorable both in the LDA+\emph{U} and GGA+\emph{U} formalisms.
Therefore, the results of NM are not presented in the following. The
dependence of the total energy (per formula unit at respective optimum
geometries) on \emph{U} for both FM and AFM phases within the LDA+\emph{U} and
GGA+\emph{U} formalisms is shown in Fig. \ref{toten}. On the whole, the total
energy of FM phase is lower than that of AFM phase no matter in LDA+\emph{U}
scheme or GGA+\emph{U} scheme. However, as shown in Fig. 1, it is clearly that
the difference can be negligible, especially as the increasing of the \emph{U}
parameter. The total-energy differences ($E_{\text{FM}}\mathtt{-}%
E_{\text{AFM}}$) within the LDA+\emph{U} at \emph{U}=4 and 6 eV are $-$0.005
eV, which is close to previous calculation \cite{Andersson} ($-$0.004 eV for
\emph{U}=4.25 eV), and $-$0.002 eV, respectively. We stress that our approach
does not include the spin-orbit coupling (SOC); a detailed explanation on this
issue can be found in Ref. \cite{Prodan1}. Both FM and AFM results will be
presented in the following analysis.

In this study, the theoretical equilibrium volume, bulk modulus \emph{B}, and
pressure derivative of the bulk modulus \emph{B$^{\prime}$} are obtained by
fitting the third-order Birch-Murnaghan equation of state (EOS) \cite{Birch}.
The elastic constants, various moduli, and Poisson's ratio $\upsilon$ are
calculated using the same method of our previous work \cite{Wang}. Note that
the bulk modulus \emph{B} obtained by these two approaches are in good
agreement, indicating that our calculations are self-consistent.

\section{results}

\subsection{Atomic and electronic structures of NpO$_{2}$}

At room temperature and zero pressure conditions, the stoichiometric neptunium
dioxide crystallize in a CaF$_{2}$-like ionic structure with space group
\emph{Fm$\bar{3}$m} (No. 225), which is the stable structure of all actinide
dioxides. Its cubic unit cell (defined by the lattice parameter $a_{0}$) is
composed of four NpO$_{2}$ formula units with the neptunium atoms and the
oxygen atoms in 4\emph{a} and 8\emph{c} sites, respectively. At higher
pressure, $\mathtt{\sim}$33 GPa experimentally, Benedict \emph{et al}.
\cite{Benedict} reported NpO$_{2}$ undergoes a phase transition to an
orthorhombic \emph{Cmcm} (No. 63) structure. In this study, we only focus our
sight on the \emph{Fm$\bar{3}$m} NpO$_{2}$.

\begin{figure}[ptb]
\begin{center}
\includegraphics[width=1.0\linewidth]{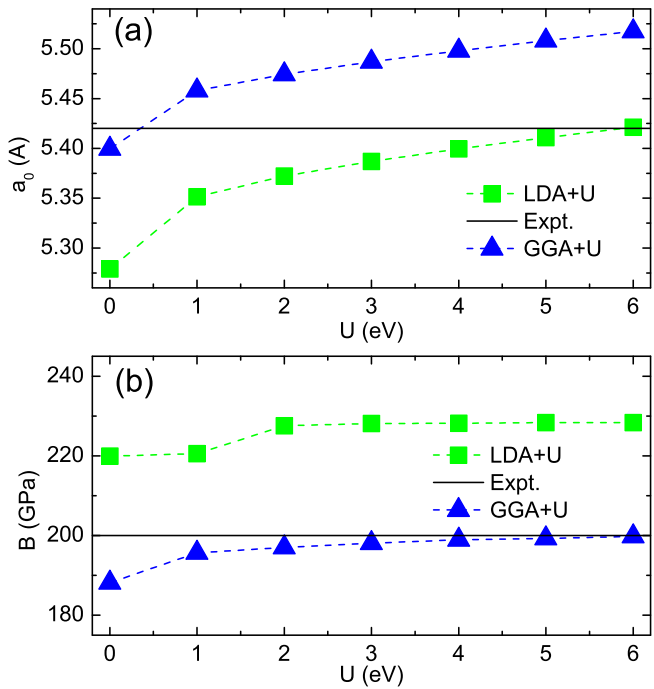}
\end{center}
\caption{(Color online) Dependence of the (a) lattice parameter and (b) bulk
modulus on \emph{U} for AFM NpO$_{2}$.}%
\label{lattice}%
\end{figure}

The present optimized lattice constant $a_{0}$ and bulk modulus \emph{B},
obtained by fitting the Murnaghan equation of state for AFM NpO$_{2}$, are
presented in Fig. \ref{lattice}. For comparison, the experimental values of
$a_{0}$ (Ref. \cite{Yamashita}) and \emph{B} (Ref. \cite{Benedict}) are also
shown. Note that the experimental value of $a_{0}$ (5.42 \AA )
\cite{Yamashita} is the fitted value at 0 K rather than at ambient condition.
In the overall view, the dependence of the lattice parameter $a_{0}$ of
NpO$_{2}$ on \emph{U} is similar to our previous study of PuO$_{2}$
\cite{SunJCP}. For the pure DFT calculations (\emph{U}=0), as shown in Fig.
\ref{lattice}(a), both LDA and GGA underestimate the lattice parameter with
respect to the experimental value. This trend is more evident for LDA due to
its overbinding character. After turning on the Hubbard \emph{U}, for the
LDA+\emph{U} approach, although the lattice parameter is still underestimated
in a wide range of \emph{U}, the calculated $a_{0}$ improves upon the pure LDA
by steadily increasing its amplitude with \emph{U}. Actually, at a typical
value \emph{U}=4 eV, the LDA+\emph{U} gives $a_{0}$=5.40 \AA \ which is very
close to the experimental value. On the other hand, the GGA+\emph{U} enlarges
the underbinding effect with increasing Hubbard \emph{U}. As a comparison, at
\emph{U}=4 eV, the GGA+\emph{U} gives $a_{0}$=5.50 \AA . Totally speaking,
both the LDA+\emph{U} and GGA+\emph{U} results of the lattice parameter for
the NpO$_{2}$ AFM phase are comparable with the experimental value at \emph{U}
around 4 eV. We have also calculated the equilibrium lattice parameter for the
FM phase for NpO$_{2}$. The tendency of $a_{0}$ with \emph{U} is similar to
that of the AFM phase. Note that the screened Coulomb hybrid functional
\cite{Prodan2} and self-interaction corrected local spin-density (SIC-LSD)
\cite{Petit} calculations predicted the lattice parameter to be 5.42 and 5.46
\AA \thinspace\ respectively.

As for the dependence of bulk modulus \emph{B} of AFM NpO$_{2}$ on \emph{U}
shown in Fig. \ref{lattice}(b), it is clear that the LDA results (220-228 GPa)
are always higher than the GGA results (188-200 GPa), which is due to the
above mentioned overbinding effect of the LDA approach. At a typical value
\emph{U}=4 eV, the LDA+\emph{U} and GGA+\emph{U} give \emph{B}=228 and 199
GPa, respectively. Apparently, the GGA+\emph{U} approach with increasing
\emph{U} gives more close values to the experimental data of \emph{B}=200 GPa
\cite{Benedict}. Note that the recent LDA+\emph{U} \cite{Andersson} and
SIC-LSD \cite{Petit} calculations predicted the bulk modulus to be 228 and 217
GPa for AFM NpO$_{2}$, respectively. Overall, comparing with the experimental
data and recent theoretical results, the accuracy of our atomic-structure
prediction for AFM NpO$_{2}$ is quite satisfactory by tuning the effective
Hubbard parameter \emph{U} in a range of 3-4 eV within the LDA/GGA+\emph{U}
approaches, which supplies the safeguard for our following study of electronic
structure and mechanical properties of NpO$_{2}$.

\begin{figure}[ptb]
\begin{center}
\includegraphics[width=1.0\linewidth]{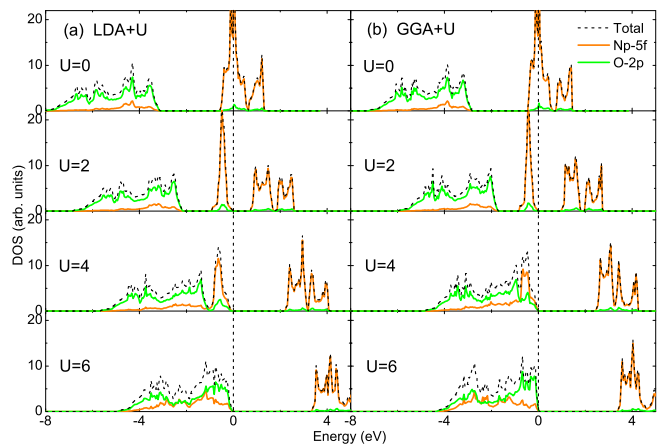}
\end{center}
\caption{(Color online) The total DOS for the NpO$_{2}$ AFM phase computed in
the (a) LDA+\emph{U} and (b) GGA+\emph{U} formalisms with four selective
values of \emph{U}. The projected DOSs for the Np 5\emph{f} and O 2\emph{p}
orbitals are also shown. The Fermi energy level is set at zero.}%
\label{DOS}%
\end{figure}

Almost all the macroscopical properties of materials, such as hardness,
elasticity, and conductivity, originate from their electronic structure
properties as well as chemical bonding nature. Therefore, a correct
description of electronic structure is necessary for further investigating
these properties. In the following, we will show the dramatic improvement
brought by the LDA/GGA+\emph{U} schemes compared to the pure ones as
describing the electronic structure properties of NpO$_{2}$. The total density
of states (DOS) as well as the projected DOS for the Np 5\emph{f} and O
2\emph{p} orbitals at four selective values of \emph{U} within
LDA/GGA+\emph{U} formalisms are shown in Fig. \ref{DOS}. Without accounting
for the on-site Coulomb repulsion (\emph{U}=0), one can see that both LDA and
GGA methods predict an incorrect metallic ground state with nonzero occupation
of Np 5\emph{f} states at \emph{E}$_{F}$. After switching on \emph{U}, as
shown in Fig. \ref{DOS}, the Np 5\emph{f} bands begin to split at
\emph{E}$_{F}$ and tend to open a Mott-type gap \emph{E$_{g}$}. The amplitude
of this insulating gap increases with increasing \emph{U }(see Fig. 4).
Overall, the LDA+\emph{U} and GGA+\emph{U} give an equivalent description of
the one-electron behaviors in a wide range of \emph{U}. At a typical value
\emph{U}=4 eV, one can see from Fig. \ref{DOS} that the occupied DOS near the
Fermi level is featured by the three well-resolved peaks. The narrow one near
$-$1.0 eV is principally Np $5f$ in character, while the other two broad peaks
respectively near $-$2.0 and $-$4.0 eV are mostly O $2p$. These well-separated
orbital peaks have been observed in the recent ultra-violet photoelectron
spectroscopy (UPS) measurement \cite{Seibert}. In addition, The width of the
Np 5\emph{f} valence band is 1 eV, consistent with the hybrid functional
calculation \cite{Prodan2} and experimental data \cite{Seibert}. The O
2\emph{p} valence band width is of 4.5 eV, also in accord with the
experimental data (4 eV) and previous hybrid functional result (4.5 eV). The
Mott gap opened at \emph{U}=4 eV is of 2.2 (2.4) eV within the
LDA(GGA)+\emph{U} scheme, see Table \ref{elastic}. Previous hybrid functional
calculations result in a larger \emph{E$_{g}$} by $\mathtt{\sim}$0.8 eV
\cite{Prodan2}. The calculated amplitude of local spin moment is $\sim$%
3.1$\mu_{B}$ (per Np atom) in both AFM and FM phases and within the two
DFT+\emph{U} schemes. On the whole, although the pure LDA and GGA fail to
depict the electronic structure, especially the insulating nature and the
occupied-state character of NpO$_{2}$, our present results show that by tuning
the effective Hubbard parameter in a reasonable range, the LDA/GGA+\emph{U}
approaches can prominently improve upon the pure LDA/GGA calculations and,
thus, can provide a satisfactory qualitative electronic structure description
comparable with the experiments. By further increasing $U$ to 6 eV, one can
see that the peak near $-$1.0 eV becomes weak and is mostly O\ 2$p$.
Meanwhile, there develops a prominent hybridization between Np 5\emph{f} and O
2\emph{p} orbitals. This picture of DOS is no longer valid since the peak near
the valence band maximum has been experimentally verified \cite{Seibert} to be
due to the Np 5\emph{f} contribution. Thus, the overestimation of $U$ will
compel Np 5\emph{f} orbitals to be even more localized, resulting in their
unphysical alignment with the O 2\emph{p} orbitals.

\begin{figure}[ptb]
\begin{center}
\includegraphics[width=1.0\linewidth]{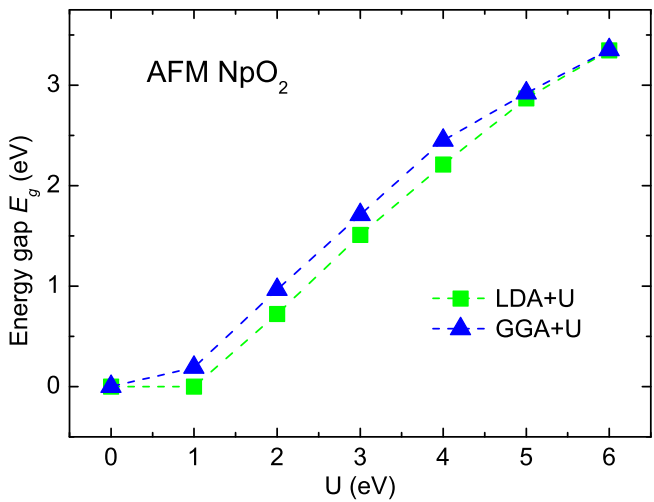}
\end{center}
\caption{(Color online) The LDA/GGA+$U$ insulating band gap for the AFM
NpO$_{2}$.}%
\end{figure}

\begin{figure}[ptb]
\begin{center}
\includegraphics[width=1.0\linewidth]{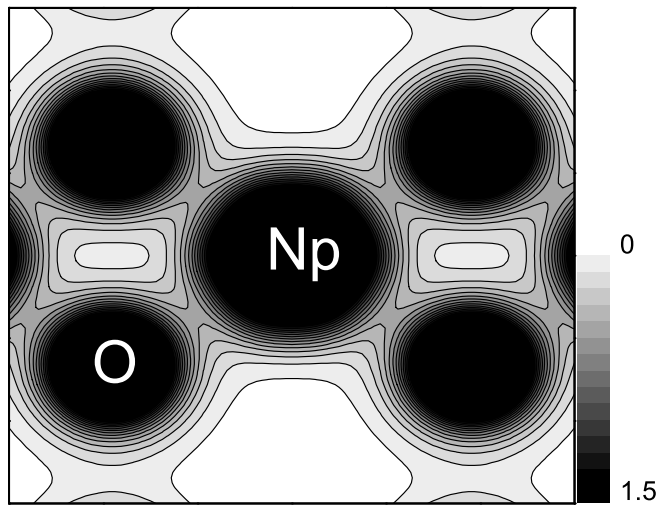}
\end{center}
\caption{Valence charge density of AFM NpO$_{2}$ in (1$\bar{1}$0) plane
computed in the LDA+\emph{U} formalisms with \emph{U}=4 eV. The contour lines
are drawn from 0.0 to 1.5 at 0.1 e/{\AA }$^{3}$ intervals.}%
\label{charge}%
\end{figure}

In order to understand the chemical bonding nature of ground state NpO$_{2}$,
we plot in Fig. \ref{charge} the valence charge density map of the (1$\bar{1}%
$0) plane. Evidently, the charge density around Np and O ions are all near
spherical distribution with slightly deformed toward the direction to their
nearest neighboring atoms. Moreover, the charge density around Np and O ions
is high while there are almost no remaining valence charges in the large
octahedral-hole interstitial region. This suggests that remarkable ionization
of neptunium and oxygen ions and significant insulating property exist for AFM
NpO$_{2}$. We have also plotted the line charge density distribution along the
nearest Np-O bond (not shown) and find that the minimum value of charge
density is 0.51 e/{\AA }$^{3}$. The distance between this minimum point and Np
ion is 1.29 \AA . Thus, the ionic radius of Np in NpO$_{2}$ can be defined as
$r_{\text{Np}}$=1.29 \AA . After subtracting this value from the Np-O bond
length (2.34 \AA ), the oxygen ionic radius $r_{\text{O}}$=1.05 \AA \ can be
obtained. The corresponding valency of Np and O in NpO$_{2}$ can be estimated
by calculating the valence charges within the spheres of ionic radii. After
integration, we find 11.386 electrons around Np and 6.233 electrons around O.
Therefore, the valency of Np and O can be represented as Np$^{3.614+}$ and
O$^{0.233-}$ respectively, which demonstrates prominent ionicity of the Np-O
bond. However, the ionicity of Np-O bond is relatively weaker when compared
with the Th-O bond in ThO$_{2}$ \cite{Wang}. On the other hand, the Np-O bond
has stronger covalency than the Th-O bond due to that the minimum value 0.51
e/{\AA }$^{3}$ of charge density along the Np-O bond in NpO$_{2}$ is
prominently larger than that along the Th-O bond (0.45 e/{\AA }$^{3}$) in
ThO$_{2}$ \cite{Wang}. This finding is consistent with the fact that the
5\emph{f} orbitals are more localized in the former in line with the
localization-delocalization trend demonstrated by the actinides series.
Through analyzing the orbital-resolved partial densities of states (PDOS) for
the AFM NpO$_{2}$, we find that the ionicity of the Np-O bond can be
attributed to the charge transfer from Np 6\emph{d} and 5\emph{f} states to O
2\emph{p} states, while the covalency of the Np-O bond is associated with the
hybridization of O $2s$ and Np $6p$ states located from $-$20.73 to $-$13.07
eV. Overall, the Np-O bond can be described as a mixture of covalent and ionic components.

\subsection{Mechanical properties of NpO$_{2}$}

\begin{table*}
\caption{Calculated elastic constants, various moduli, pressure
derivative of the bulk modulus \emph{B$^{^{\prime}}$}, Poisson's
ratio $\upsilon$, spin moments ($\mu_{mag.}$), and insulating band
gap (\emph{E$_{g}$}) for \emph{Fm$\bar{3}$m} NpO$_{2}$ at 0 GPa. For
comparison, experimental values
from Ref. \cite{Benedict} are also listed.}%
\label{elastic}
\begin{ruledtabular}
\begin{tabular}{cccccccccccccccc}
Magnetism&Method&\emph{C$_{11}$}&\emph{C$_{22}$}&\emph{C$_{44}$}&\emph{B}&\emph{B$^{'}$}&\emph{G}&\emph{E}&$\upsilon$&$\mu_{mag.}$&E$_{g}$\\
&&(GPa)&(GPa)&(GPa)&(GPa)&&(GPa)&(GPa)&&($\mu_{B}$)&(eV)\\
\hline
AFM&LDA+\emph{U} (\emph{U}=0)&405.1 & 128.8  & 78.9 &221&4.1&98.9&258.2&0.305&2.678&0.0\\
&LDA+\emph{U} (\emph{U}=4)&399.5 & 145.5  & 72.9 &230&4.5&91.2&241.6&0.325&3.074&2.2\\
&GGA+\emph{U} (\emph{U}=0)&358.7 & 105.2  & 53.5 &190&4.2&76.2&201.7&0.323&2.903&0.0\\
&GGA+\emph{U} (\emph{U}=4)&363.6 & 118.8  & 57.4 &200&4.3&78.1&207.5&0.327&3.092&2.4\\
&Expt.& &  &  &200&3.8&&&&&\\
FM&LDA+\emph{U} (\emph{U}=0)&415.0 &148.7  & 79.3 &237&4.5&97.7&257.8&0.319&3.054&0.0\\
&LDA+\emph{U} (\emph{U}=4)&403.6 &143.4  &73.7  &230&4.2&92.7&245.2&0.323&3.090&2.1\\
&GGA+\emph{U} (\emph{U}=0)&372.7 &116.8  &70.9  &202&4.4&90.0&235.1&0.306&3.104&0.0\\
&GGA+\emph{U} (\emph{U}=4)&365.8 &117.8  &57.5  &200&4.5&78.6&208.6&0.327&3.103&2.4\\
\end{tabular}
\end{ruledtabular}
\end{table*}

Our calculated elastic constants, various moduli, pressure derivative of the
bulk modulus \emph{B$^{^{\prime}}$}, and Poisson's ratio $\upsilon$ for
\emph{Fm$\bar{3}$m} NpO$_{2}$ at 0 GPa are collected in Table \ref{elastic}.
Here, the moduli and Poisson's ratio are deduced from the calculated elastic
constants using the same scheme as that in our previous study on isostructural
ThO$_{2}$ \cite{Wang}. Each derived bulk modulus \emph{B} turns out to be very
close to that obtained by fitting Murnaghan equation of state. This indicates
that our calculations are consistent and reliable. At a typical value of
\emph{U}=4 eV, LDA+\emph{U} gives bulk modulus of 230 GPa for both AFM and FM
NpO$_{2}$, while GGA+\emph{U} gives 200 GPa. Obviously, GGA+\emph{U} approach
gives more closer value than LDA+\emph{U} with respect to the experimental
value. Overall, the bulk modulus of NpO$_{2}$ is bigger than that of ThO$_{2}$
\cite{Wang}. This originates from relative higher valence electron density and
shorter bond distances of NpO$_{2}$ than those of ThO$_{2}$.

In addition, the hardness of NpO$_{2}$ is investigated by using the approach
raised by Simunek \emph{et al.} \cite{SimunekPRL}. In the case of two atoms 1
and 2 forming one bond of strength \emph{s$_{12}$} in a unit cell of volume
$\Omega$, the expression for hardness has the form \cite{SimunekPRL}
\begin{equation}
H=(C/\Omega)b_{12}s_{12}e^{-\sigma\!f_{2}},
\end{equation}
where
\begin{equation}
s_{12}=\sqrt{(e_{1}e_{2})}/(n_{1}n_{2}d_{12}), e_{i}=Z_{i}/r_{i}%
\end{equation}
and
\begin{equation}
f_{2}=(\frac{e_{1}-e_{2}}{e_{1}+e_{2}})^{2}=1-[\sqrt{(e_{1}e_{2})}%
/(e_{1}+e_{2})]^{2}%
\end{equation}
are the strength and ionicity of the chemical bond, respectively, and
\emph{d$_{12}$} is the interatomic distance; \emph{C}=1550 and $\sigma$=4 are
constants. The radius \emph{r$_{i}$} is chosen to make sure that the sphere
centered at atoms \emph{i} in a crystal contains exactly the valence
electronic charge \emph{Z$_{i}$}. For fluorite structure NpO$_{2}$,
\emph{b$_{12}$}=32 counts the interatomic bonds between atoms Np (1) and O (2)
in the unit cell, \emph{n$_{1}$}=8 and \emph{n$_{2}$}=4 are coordination
numbers of atom Np and O, respectively, \emph{r$_{1}$}=1.715 ({\AA }) and
\emph{r$_{2}$}=1.000 ({\AA }) are the atomic radii for Np and O atoms,
respectively, \emph{Z$_{1}$}=15 and \emph{Z$_{2}$}=6 are valence charge for Np
and O atoms, respectively, \emph{d$_{12}$}=2.34 ({\AA }) is the interatomic
distance, and $\Omega$=157.42 ({\AA }$^{3}$) is the volume of unit cell. Using
Eqs. (2)-(4), we obtain \emph{s$_{12}$}=0.0967 and \emph{f$_{2}$}=0.0347. The
hardness of NpO$_{2}$ at its ground-state fluorite structure is thus given by
\emph{H}=26.5 (GPa). This indicates that the fluorite NpO$_{2}$ is a hard
material and approaches to a superhard material (hardness $>$ 40 GPa). The
high hardness of this crystal can be understood from the dense crystal
structure, which results in high valence electron density and short bond
distances. We notice that the hardness of NpO$_{2}$ is higher than that of
ThO$_{2}$ (22.4 GPa) \cite{Wang}. This also can be attributed to the higher
electron density and shorter bond distance of NpO$_{2}$ compared with
ThO$_{2}$.

\subsection{Phonon dispersion curve of NpO$_{2}$}

Before phonon dispersion calculation, we have calculated the Born effective
charges and dielectric constants of NpO$_{2}$ because of their critical
importance to correct the LO-TO splitting. The Born effective charge $Z^{\ast
}$ is a measurement of the change in electronic polarization due to the ionic
displacements. The form of $Z^{\ast}$ for a particular atom depends on the
atomic position symmetry. For fluorite NpO$_{2}$, the effective charge tensor
for both Np and O that occupy the 4\emph{a} and 8\emph{c} Wyckhoff positions
respectively are isotropic. After calculation, the Born effective charges of
Np and O ions for AFM (FM) NpO$_{2}$ are found to be $Z_{\mathrm{{Ce}}}^{\ast
}$=5.24 (5.26) and $Z_{\mathrm{{O}}}^{\ast}$=$-$2.62 ($-$2.63), respectively,
within LDA+\emph{U} formalism with the choice of $U$=$4.0$ eV. In addition,
the macroscopic static dielectric tensor is also isotropic and our computed
value of dielectric constant $\varepsilon_{\infty}$ is 5.58 for the AFM phase
and 5.64 for the FM phase.

\begin{figure}[ptb]
\begin{center}
\includegraphics[width=1.0\linewidth]{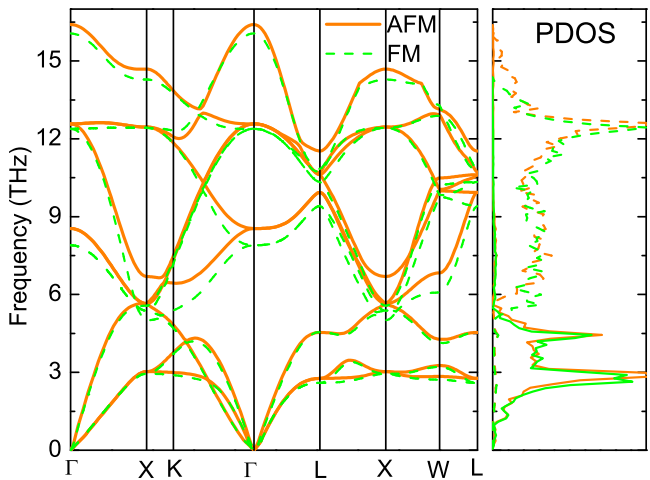}
\end{center}
\caption{(Color online) Calculated phonon dispersion curves (left
panel) and corresponding PDOS ( right panel) for NpO$_{2}$. Both the
AFM and FM results are calculated within LDA+\emph{U} formalism with
\emph{U}=4 eV. In right panel, the orange and green lines stand for
AFM and FM results, respectively, while solid and
dashed lines are used to distinguish the PDOS of Np and O.}%
\label{phonon}%
\end{figure}

Employing the Hellmann-Feynman theorem and the direct method, we have
calculated the phonon curves along some high-symmetry directions in the
Brillouin zone (BZ), together with the phonon density of states. For the
phonon dispersion calculation, we use the 2$\times$2$\times$2 fcc supercell
containing 96 atoms and the 4$\times$4$\times$4 Monkhorst-Pack \emph{k}-point
mesh for the BZ integration. In order to calculate the Hellmann-Feynman
forces, we displace four atoms (two Np and two O atoms) from their equilibrium
positions and the amplitude of all the displacements is 0.03 \AA . The
calculated phonon dispersion curves along the $\Gamma$$-$$X$$-$$K$$-$$\Gamma
$$-$$L$$-$$X$$-$$W$$-$$L$ directions are displayed in Fig. \ref{phonon} for
both FM and AFM phases of NpO$_{2}$.

To our knowledge, no experimental phonon frequency results have been published
for NpO$_{2}$. In the CaF$_{2}$-structure primitive cell, there are only three
atoms (one Np and two O atoms). Therefore, nine phonon modes exist in the
dispersion relations. One can see that the gap between the optic modes and the
acoustic branches is not so evident. But the LO-TO splitting at $\Gamma$ point
is very significant since the inclusion of polarization effects. Due to the
fact that neptunium is heavier than oxygen atom, the vibration frequency of
neptunium atom is lower than that of oxygen atom. Therefore, the phonon
density of states for NpO$_{2}$ can be viewed as two parts. One is the part
lower than 5.7 (5.0) THz for AFM (FM) NpO$_{2}$ where the main contribution
comes from the neptunium sublattice, while the other part higher than 5.7
(5.0) THz are dominated by the dynamics of the light oxygen atoms. In
addition, one can see from Fig. \ref{phonon} that the difference in the phonon
dispersion between the FM and AFM phases of NpO$_{2}$ is neglectable in most
regions of the BZ, except for one TO branch, which is a little lower for the
FM phase than for the AFM phase.

\section{CONCLUSION}

In conclusion, we have studied the structural, electronic, mechanical, and
thermodynamic properties of NpO$_{2}$ within the LDA+\emph{U} and GGA+\emph{U}
formalisms. The atomic structure, including lattice parameters and bulk
modulus, and the one-electron behaviors of 5\emph{f} state have been
systematically investigated as a function of the effective on-site Coulomb
repulsion parameter \emph{U}. By choosing the Hubbard \emph{U} parameter
around 4 eV within the LDA/GGA+\emph{U} approaches, most of our calculated
results are in good agreement with the experiments. Based on this, we have
further investigated the electronic structures of NpO$_{2}$ and the Np-O bond
nature. The insulating ground state can be well reproduced and the Np-O bond
is found to be of a mixture of both the ionic and covalent characters. As for
the mechanical property, we find that GGA+\emph{U} approach gives more closer
value of bulk modulus than LDA+\emph{U} with respect to the experimental value
and the hardness of NpO$_{2}$ is calculated to be 26.5 GPa. For the
thermodynamic analysis, the Born effective charges and dielectric constants as
well as the phonon dispersion curves have also been presented.

\begin{acknowledgments}
This work was supported by the Foundations for Development of Science and
Technology of China Academy of Engineering Physics under Grant No. 2009B0301037.
\end{acknowledgments}

\end{document}